\documentclass[aps,prd,twocolumn,groupedaddress,showpacs]{revtex4}
\usepackage{epsf,epsfig,graphics,graphicx}
\begin{document}

\title{Production of high stellar-mass primordial black holes in trapped inflation}

\author{Shu-Lin Cheng$^1$}
\author{Wolung Lee$^1$}
\author{Kin-Wang Ng$^{2,3}$}

\affiliation{
$^1$Department of Physics, National Taiwan Normal University,
Taipei 11677, Taiwan\\
$^2$Institute of Physics, Academia Sinica, Taipei 11529, Taiwan\\
$^3$Institute of Astronomy and Astrophysics, Academia Sinica, Taipei 11529, Taiwan}

\begin{abstract}
Trapped inflation has been proposed to provide a successful inflation with a steep potential. We discuss the formation of primordial black holes in the trapped inflationary scenario. We show that primordial black holes are naturally produced during inflation with a steep trapping potential. In particular, we have given a recipe for an inflaton potential with which particle production can induce large non-Gaussian curvature perturbation that leads to the formation of high stellar-mass primordial black holes. These primordial black holes could be dark matter observed by the LIGO detectors through a binary black-hole merger. At the end, we have given an attempt to realize the required inflaton potential in the axion monodromy inflation, and discussed the gravitational waves sourced by the particle production.
\end{abstract}

\maketitle

The LIGO detectors have firstly observed gravitational waves predicted in general relativity~\cite{ligo}. The discovery opens up a new era of gravity-wave astronomy and cosmology. The detected gravity-wave source GW150914 provides an evidence for the existence of high stellar-mass binary black-hole (BBH) systems with masses $29 M_\odot$ and $36 M_\odot$ that inspiral and merge within the age of the Universe, with an inferred merger rate of $2-400 {\rm Gpc}^{-3} {\rm yr}^{-1}$~\cite{ligorate}. So far, neither electromagnetic nor neutrino counterparts have been confirmed, implying that gas accretion onto the BH binary is insignificant at the time of merging.

In astrophysics, the observation of GW150914 has constrained the theory of stellar-mass BBH formation. The merger rate inferred from the observation of GW150914 has excluded the formation models with low merger rates ($\lesssim 1 {\rm Gpc}^{-3} {\rm yr}^{-1}$)~\cite{ligobh}. 
The low measured redshift ($z\simeq 0.1$) of GW150914 and the low inferred metallicity of the stellar progenitor imply either BBH formation in a local dwarf galaxy followed by a prompt merger, or formation at high redshift with a long delayed merger. There remain two prevailing formation categories for BBH mergers: dynamical gravitational interactions between BHs and stars in dense globular clusters (Ref.~\cite{dynamical} and references therein) and isolated stellar binaries in galactic fields (Ref.~\cite{field} and references therein).

Recently, inspired by LIGO results, the authors in Ref.~\cite{bird} have considered a possibility that the BH binary associated with GW150914 may be of primordial origin, different from more traditional astrophysical sources - the two BHs are primordial black holes (PBHs) formed in the early Universe. Interestingly, massive PBHs being non-relativistic and weakly interacting gravitationally may constitute the cold dark matter, if they are formed prior to the big bang nucleosynthesis. The estimated merger rate for these PBHs, if they are clustered in compact sub-halos, can span a range that overlaps the BH merger rate inferred from GW150914~\cite{bird,clesse,sasaki}. As such, GW150914 can be considered as an indirect signal of dark matter coming from the PBH sub-halo that is a conversion of a few $M_\odot$ dark matter into gravitational waves in a massive PHB coalescence. This may explain why GW150914 has no optical counterpart, because sub-halo PBHs deplete gas and quench star formation. Furthermore, PBH sub-halos should manifest as faint dwarf galaxies, possibly alleviating the missing satellite and too-big-to-fail problems. 

Although the PBH alternative for GW150914 seems plausible, many challenging questions such as their formation and subsequent evolution into sub-halos, the mass fraction in dark matter, and the distinction between PBHs and astrophysical BHs, remain to be answered. In fact, there are astrophysical and cosmological constraints that have already excluded PBHs as the main component of dark matter over a vast range of PBH masses~\cite{carr,green}. Gravitational microlensing surveys have excluded masses $\lesssim 10 M_\odot$, while higher masses ($\gtrsim 100 M_\odot$) would disrupt galactic wide binary stars. Nevertheless, the constraints on the PBH mass between $10 M_\odot$ and $100 M_\odot$, derived from cosmic microwave background (CMB) observations, are controversial, because they require modeling of several complex physical processes inherent with significant uncertainties. The existence of high stellar-mass PBHs not only solves the dark matter problem, but also provides an important test on the theory of PBH formation in the early Universe.

There are various scenarios for forming PBHs over a wide range of masses in the early Universe~\cite{green}. Apart from those spontaneously formed in phase transitions, PBHs can be arisen from the collapse of horizon-sized large matter inhomogeneities. This matter density perturbation may originate in quantum fluctuations during inflation that re-enter the horizon in the subsequent expanding Universe. 
If a seed to form a PBH is created during inflation, the mass of the PBH can be estimated as follows. The energy contained within the comoving seed volume that leaves the horizon $N$ e-foldings before the end of inflation is governed by
%eq.1
\begin{equation}
{4\over 3}\pi H^{-3} \rho\,e^{3N}~~~{\rm with}~ \quad \rho=3H^2 M_p^2,
\end{equation}
where $H$ is the Hubble constant during inflation and $M_p$ the reduced Planck mass. After the inflation has ended, the Universe reheats and becomes radiation-dominated. Let $a_0=1$ be the scale factor at the onset of inflation, which will eventually span $N_0$ e-foldings over the entire inflationary course. Subsequently, in terms of the conformal time $\eta$ defined by $dt=a(\eta) d\eta$, a flat universe would have expanded up to a size of $a(\eta)=H e^{2N_0} \eta$ in the radiation-dominated epoch. The comoving volume re-enters the horizon when its scale $k=H e^{N_0-N}$ satisfies the condition $k\eta\sim 1$, i.e., when $a=e^{N_0+N}$ or the temperature of the thermal bath is red-shifted by a factor of $e^N$~\cite{ng}. Therefore, the mass of the PBH presumably formed at this time is 
%eq.2
\begin{equation}
M_{\rm BH}=\frac{4 \pi M_p^2}{H}\,e^{2N}=2.74\times 10^{-38} e^{2N} \left(\frac{M_p}{H}\right) M_\odot.
\label{mbh}
\end{equation}

In single-field slow-roll inflation models, the matter density perturbation is generally well below the threshold to form PBHs though they can be formed at rare density peaks. Modifications of the inflation potential to achieve blue-tilted matter power spectra or running spectral indices may lead to large density perturbation at the end of inflation~\cite{drees}. When inflaton is coupled to quantum fields, copious quanta production can induce large perturbation at small scales too~\cite{lin,linde,cheng,bugaev}. However, the resulting PBH masses in most of these models are many orders of magnitude below $M_\odot$. To boost the PBH mass into astrophysical and even cosmological mass scales, several scenarios involving multi-field inflation have been proposed, such as the hybrid inflation~\cite{hybrid}, the double inflation~\cite{double}, and the curvaton models~\cite{curvaton}, in which small-scale density perturbation can be inflated to a scale ranging from the size of a stellar-mass PBH to a supermassive PBH. The astrophysical and cosmological bounds on PBHs have been used to exclude or constrain all these models. In this paper, we propose a mechanism for the production of high stellar-mass PBHs in a single-field trapped inflation~\cite{kofman04,green09,anber10,trap}. 

Trapped inflation has been proposed to provide a successful inflation with a steep potential. Due to a coupling of the inflaton with other massless particles, the inflaton in the trapped inflation rolls slowly down a steep potential by dumping its kinetic energy into particle production~\cite{kofman04,green09,anber10}. As shown in Ref.~\cite{green09}, a viable trapped inflation model can be achieved even on a potential which is too steep for slow-roll inflation; in addition, the inflaton fluctuations induced by the backreaction of the particle production dominate over the vacuum inflaton fluctuations. In Ref.~\cite{anber10}, the authors considered an axion-like inflaton coupled to massless gauge fields through a Chern-Simons term. They found that the inflation with a steep potential is effectively a slow-roll inflation but unfortunately the induced inflaton fluctuations become too large to satisfy CMB constraints. Here we adopt the main idea of trapped inflation but we modify it in the following manner. In order to explain the CMB anisotropy data, we still use a flat potential for driving a standard slow-roll inflation for the first $15$ e-foldings after the beginning of inflation. After that, the inflaton slides into a steeper potential well, experiencing a trapped type of inflation. Large inflaton fluctuations will be produced by profuse particle production in this trapped period until the inflaton exits the potential well and rolls towards the end of inflation. These large inflaton fluctuations during the trapped inflation are the seed for the formation of PBHs when the fluctuations exit and 
re-enter the horizon. 

To achieve our goal, we consider a modified version of the trapped inflation driven by a pseudoscalar $\varphi$ that couples to a $U(1)$ gauge field $A_\mu$:
%eq.3
\begin{eqnarray}
  \mathcal{S} &=&  \int d^4 x \sqrt{-g} \left[ \frac{M_p^2}{2} \, R  -\frac{1}{2}\partial_\mu\varphi \partial^\mu\varphi- V(\varphi) \right.\nonumber\\
&& \left. - \frac{1}{4}F^{\mu\nu}F_{\mu\nu} - \frac{\alpha}{4 f} \varphi  \, \tilde{F}^{\mu\nu} \, F_{\mu\nu} \right],
\label{action}
\end{eqnarray}
where $R$ is the curvature scalar, $V(\varphi)$ is the inflaton potential, $\alpha$ is a coupling constant, $f$ is an energy scale, $F_{\mu \nu} = \partial_\mu A_\nu - \partial_\nu A_\mu$ is the field strength tensor, and $\tilde{F}^{\mu\nu} = \frac{1}{2}\epsilon^{\mu\nu\alpha\beta} F_{\alpha\beta}/{\sqrt{-g}}$ is its dual. The potential $V(\varphi)$ is a standard flat potential superimposed with a shallow potential well.

To calculate the production of gauge quanta, we separate the inflaton into a mean field and its fluctuations:
%eq.4
\begin{equation}
  \varphi = \phi (\eta) + \delta \varphi (\eta,{\vec x}).
\end{equation}
Under the temporal gauge, $A_{\mu} = (0, {\vec A})$, we decompose ${\vec A}(\eta,{\vec x})$ into its right and left circularly polarized Fourier modes, $A_\pm (\eta,{\vec k})$,
whose equation of motion is then given by
%eq.5
\begin{equation}
  \left[ \frac{d^2}{d\eta^2} + k^2 \mp 2aH k\xi\right] A_{\pm}(\eta,k) = 0, \;\; \xi \equiv \frac{\alpha}{2 f H}\frac{d\phi}{dt}\,.
\label{photoneom}
\end{equation}
This equation implies that either one of the two modes, satisfying the condition $k/(aH)< 2|\xi|$, becomes unstable and grows exponentially.
The energy density and the interaction term of the produced gauge quanta are given by the vacuum expectation values of the electric and magnetic fields, respectively,
%eq.6~7
\begin{eqnarray}
\frac{1}{2}\langle \vec{E}^2+\vec{B}^2 \rangle\hspace{-1mm}&=&\hspace{-1mm}\int\frac{dk\,k^2}{4 \pi^2 a^4} \sum_{\lambda=\pm}\left( \left\vert \frac{dA_\lambda}{d\eta} \right\vert^2 + k^2 \vert A_\lambda \vert^2 \right)\hspace{-1mm}, \\
\langle \vec{E} \cdot \vec{B} \rangle &=& -\int\frac{dk\,k^3}{4 \pi^2 a^4} \frac{d}{d \eta} \left(\vert A_+ \vert^2-\vert A_- \vert^2\right).
\end{eqnarray}
On the other hand, the production of gauge quanta gives rise to a backreaction on the background. The background evolution is therefore governed by
%eq.8~9
\begin{eqnarray}
  &&  \frac{d^2\phi}{dt^2} + 3 H \frac{d\phi}{dt} +\frac{dV}{d\phi} = \frac{\alpha}{f} \langle \vec{E}\cdot \vec{B} \rangle, \label{inflatoneom} \\
  && 3 H^2 = \frac{1}{M_p^2} \left[ \frac{1}{2}\left(\frac{d\phi}{dt}\right)^2 + V(\phi) + \frac{1}{2} \langle \vec{E}^2 + \vec{B}^2 \rangle \right].
\label{Heom}
\end{eqnarray}
Then, the generation of inflaton fluctuations during inflation can be calculated by
%eq.10
\begin{eqnarray}%&&
\left[ \frac{\partial^2}{\partial t^2} +3 \beta H \frac{\partial}{\partial t} -\frac{{\vec\nabla}^2}{a^2} \hspace{-2mm} \right. &+& \hspace{-2mm} \left. \frac{d^2V}{d\phi^2} \right] \delta\varphi(t,{\vec x}) \nonumber \\ 
&=& {\alpha\over f}\left(  \vec{E}\cdot\vec{B} - \langle \vec{E}\cdot\vec{B} \rangle \right), 
\end{eqnarray}
where the dissipation effect is depicted as~\cite{anber2,barnaby2}
%eq.11
\begin{equation}
\beta\equiv 1-2\pi\xi \frac{\alpha}{f} \frac{\langle \vec{E}\cdot\vec{B} \rangle}{3H(d\phi/dt)}.
\end{equation}
It was shown that the solution to this equation can be well approximated by~\cite{barnaby2,linde}
%eq.12
\begin{equation}
\delta\varphi= {\alpha\over{3\beta f H^2}}\left(  \vec{E}\cdot\vec{B} - \langle \vec{E}\cdot\vec{B} \rangle\right),
\end{equation}
which leads to a contribution to the power spectrum of the curvature perturbation given by
%eq.13
\begin{equation}
\Delta_\zeta^2(k)=\langle\zeta(x)^2\rangle=\frac{H^2 \langle\delta\varphi^2\rangle}{(d\phi/dt)^2}
=\left[\frac{\alpha\langle \vec{E}\cdot\vec{B} \rangle}{3\beta f H(d\phi/dt)}\right]^2.
\label{Deltazeta}
\end{equation}

\begin{figure}[htp]
\centering
\includegraphics[width=0.4\textwidth]{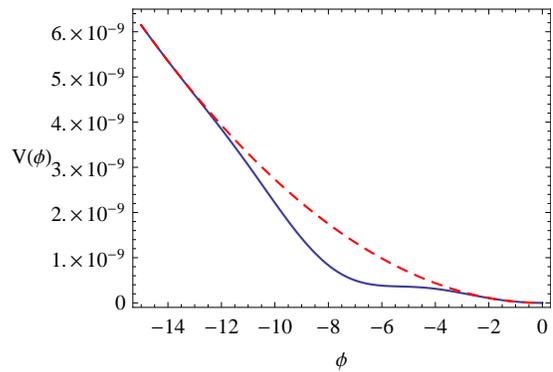}
\caption{Inflaton potential $V(\phi)$ with the dashed line denoting the quadratic term, $m^2\phi^2/2$. All dynamical variables in this figure and in the following figures are
rescaled by the reduced Planck mass, $M_p=2.435\times 10^{18}$ GeV. }
\label{potential}
\end{figure}

\begin{figure}[htp]
\centering
\includegraphics[width=0.4\textwidth]{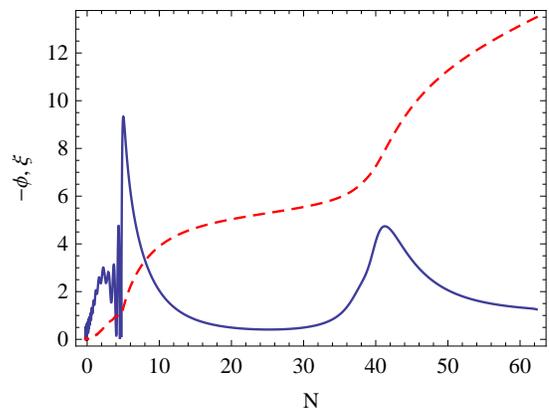}
\caption{Evolutions of $\phi$ (dashed line) and $\xi$ (solid line) as functions of e-foldings $N$ before the end of inflation. Inflation starts at $N_0\simeq 61$.}
\label{phixi}
\end{figure}

\begin{figure}[htp]
\centering
\includegraphics[width=0.4\textwidth]{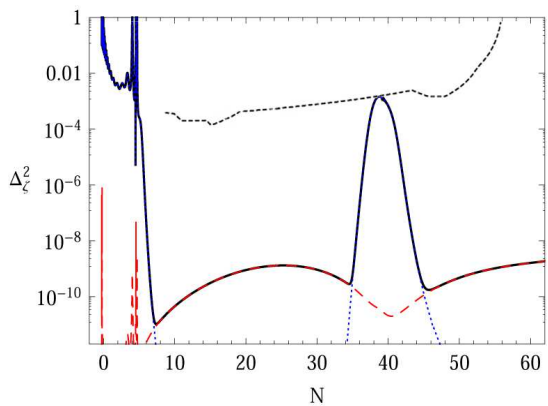}
\caption{Solid line is the total power spectrum of the curvature perturbation. The contribution induced by photon production is denoted by the dotted line and the vacuum contribution by the dashed line. The e-folding $N$ denotes the time when the $k$-mode leaves the horizon. The primordial black hole bound is the short-dashed line. }
\label{power}
\end{figure}

Following the numerical scheme in our previous paper~\cite{cheng}, we compute the power spectrum~(\ref{Deltazeta}) by solving numerically the coupled differential equations of motion for inflaton and photon mode functions in Eqs.~(\ref{photoneom}), (\ref{inflatoneom}), and (\ref{Heom}) from the onset of inflation to the end of inflation, included self-consistently with the backreaction due to photon production. We employ a modified quadratic potential,
\begin{equation}
V(\varphi)={m^2\over 2}\left[\varphi^2 +a M_p^2 \sin^n\left(\frac{\varphi}{bM_p}\right)\right],
\label{recipe}
\end{equation}
where $m=1.8\times 10^{13}$ GeV, $a=-33.6$, $b=5$, and $n=6$. Here we rescale all dynamical variables in terms of the reduced Planck mass, $M_p=2.435\times 10^{18}$ GeV. Hence, $m=7.39\times 10^{-6}$. We set $\alpha=17$, $f=1$, and $\phi_0=-13.5$ and $(d\phi/dt)_0=6.02\times 10^{-6}$ respectively for the initial position and speed of the inflaton. The number of e-foldings after the beginning of inflation is defined by $\int_0^t H(t') dt'$. Note that $a_0=1$ and we find that $N_0\simeq 61$. Figure~\ref{potential} shows the inflaton potential $V(\phi)$. The evolutions of $\phi$ and $\xi$ as functions of the e-foldings $N$ before the end of inflation are shown in Fig.~\ref{phixi}.  Figure~\ref{power} depicts the total power spectrum of the curvature perturbation, which is the sum of the induced power in Eq.~(\ref{Deltazeta}) and the vacuum contribution given by $H^4/(4\pi^2(d\phi/dt)^2)$.

\begin{figure}[htp]
\centering
\includegraphics[width=0.4\textwidth]{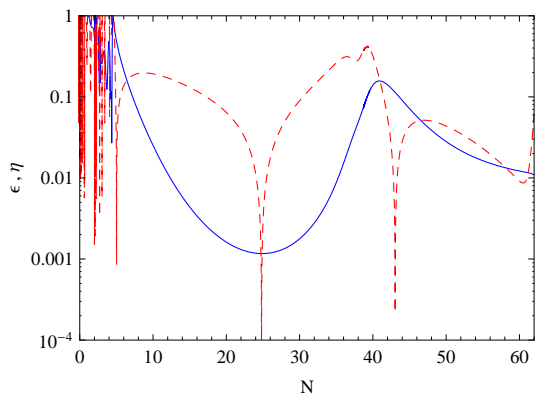}
\caption{Evolutions of the slow-roll parameters $\epsilon$ (solid line) and $\eta$ (dashed line) as functions of e-foldings $N$.}
\label{slowroll}
\end{figure}

The $V(\phi)$ is a quadratic potential modified by a shallow well. In the first $10$ e-foldings, the inflaton rolls slowly in the quadratic part of the potential in a manner as that in the chaotic inflation~\cite{chaotic}. As in standard slow-roll inflation, the power spectrum of the curvature perturbation is dominated by vacuum fluctuations, being consistent with the amplitude $\Delta_\zeta^2\simeq 2\times 10^{-9}$ at large scales measured by Planck mission~\cite{planck}. 
The slow-roll parameters, $\epsilon\equiv -(dH/dt)/H^2$ and $\eta\equiv (d^2\phi/dt^2)/(Hd\phi/dt)$, are at a level of $0.01-0.03$ in the first $10$ e-foldings, as shown in Fig.~\ref{slowroll}. The predicted spectral index of the curvature perturbation power spectrum is then given by $n_s\simeq 1-6\epsilon+2\eta$, which is consistent with the Planck measurement, $n_s\simeq 0.97$~\cite{planck}. This can also be seen as a red-tilted power spectrum in Fig.~\ref{power}.

When the inflaton enters the steeper slope of the well, its speed increases and so does $\xi$. This results in a huge production of photons and a largely induced inflaton fluctuations at $N\simeq 40$. Meanwhile, the backreaction due to the photon production slows down the inflaton motion; in particular, it drags the inflaton at $\phi\simeq -5$ for about $20$ e-foldings and thus prolongs the duration of inflation. Near the end of inflation, the inflaton speeds up again and hence produces a large amount of photons and inflaton fluctuations. Finally, inflation ends up with a rather complicated dissipation-fluctuation process. However, because of the strong backreaction to the inflaton motion, the inflaton gradually reduces to zero field value and inflation ends gracefully without undergoing preheating or typical perturbative reheating. During the full course of a hybrid type of slow-roll and trapped inflation, the vacuum energy is being drained by the production of photons and converted into radiation. Interestingly, for a sufficiently large coupling $\alpha$, the production of photons can sustain a nearly steady thermal bath during inflation and exhaust the vacuum energy to end the inflation gracefully~\cite{cheng}.

Now we turn to the production of PBHs. In Eq.~(\ref{mbh}), by taking $H=5.5\times 10^{-6}$, $M_{\rm BH}\simeq 5\times 10^{-33} e^{2N} M_\odot \simeq 10 e^{2N}$g. 
In Fig.~\ref{power}, the short-dashed line is the upper bound on the power spectrum coming from the non-detection of PBHs~\cite{linde}. Note that the bound ends at $N\simeq 8$, corresponding to $M_{\rm BH}\simeq 10^8$g, because the constraints on the PBH mass $M_{\rm BH} < 10^8$g are strongly model dependent. 
The power spectrum induced by the photon production shows a prominent peak at around $N=40$ that saturates the PBH bound at $N\simeq38.5$, so PBHs of mass about 
$14 M_\odot$ can be significantly formed, with the black hole fraction $\beta(M)\sim 10^{-7}$, which gives the fraction of dark matter in PBHs $f(M)\sim 0.1$~\cite{carr,linde}. Near the end of inflation the power spectrum shows high spikes where PBHs of much smaller masses are likely to be copiously formed as well. We warn that in the present consideration the power spectrum is no longer valid when its value becomes near unity. In this situation, one must take into account the effects of gravitation, which could presumably dampen the spikes to a consistent level~\cite{fugita}.

We have given a novel model associated with trapped inflation in which high stellar-mass PBHs can be formed. Although we have designed a modified quadratic potential to achieve this, we anticipate that many similar potentials can fulfill the requirements. To realize our model, we have studied a case in the context of axion monodromy inflation~\cite{silverstein}. The potential in single-field axion monodromy inflation has a monomial form with superimposed modulations whose size is model-dependent, given by~\cite{flauger}
\begin{equation}
V(\varphi)=V_0+\mu^{4-p} \varphi^p +\Lambda(\varphi)^4 \cos\left[\frac{\varphi}{f(\varphi)}+\gamma_0\right],
\end{equation}
where $V_0$, $\mu$, $p$, and $\gamma_0$ are constants. The modulation contains the energy scale $\Lambda$ and the axion decay constant $f$, which are in general functions of $\varphi$. We have tuned the model parameters to obtain the potential,
\begin{equation}
V(\varphi)={1\over 2}m^2\varphi^2 +a m^2 M_p^2\left[1+\cos\left(\frac{\varphi}{bM_p}+\pi\right)\right],
\end{equation}
where $a=2.24$ and $b=0.712$. With $\alpha=18.4$, $\phi_0=-10.5$, and $(d\phi/dt)_0=6.02\times 10^{-6}$, this potential gives similar results to the potential~(\ref{recipe}) for the time evolutions of $\phi$ and $\xi$ as well as the power spectrum of the curvature perturbation as shown in Fig.~\ref{phixi} and Fig.~\ref{power}, respectively. It would be interesting to perform a systematic search for the parameter space in the modulations that allow for the formation of high stellar-mass PBHs. Also, we can use existing astrophysical and cosmological constraints on a wide range of PBH masses to assess axion couplings and modulations in axion monodromy inflation.

\begin{figure}[htp]
\centering
\includegraphics[width=0.4\textwidth]{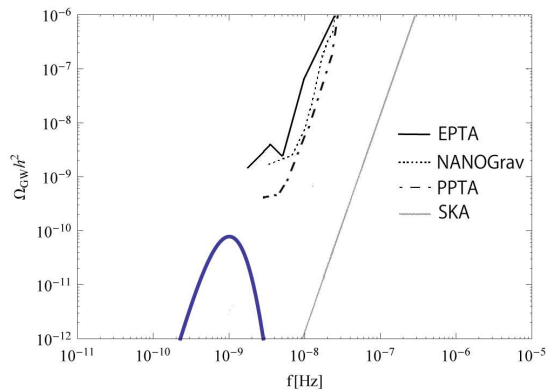}
\caption{Present spectral energy density of the gravitational waves associated with the production of $14 M_\odot$ PBHs. Also shown are the upper limits set by the pulsar timing array experiments EPTA,  NANOGrav, and PPTA, as well as the projected SKA sensitivity.}
\label{pbhgw}
\end{figure}

The produced photons also source tensor metric perturbation or gravitational waves, 
\begin{equation}
\left[\frac{\partial^2}{\partial \eta^2} + {2\over a}\frac{da}{d\eta} \frac{\partial}{\partial \eta} -{\vec\nabla}^2\right] h_{ij}
= \frac{2a^2}{M_p^2}\left( -E_i E_j - B_i B_j \right)^{TT}, 
\end{equation}
where $TT$ denotes the transverse and traceless projection of the spatial components of the energy-momentum tensor of the gauge field. This equation can be simultaneously solved with the numerical solution for the photon mode function in Eq.~(\ref{photoneom}). Instead of pursuing this computationally heavy task,  we estimate the tensor perturbation by using the results based on analytic photon mode functions found in Refs.~\cite{NGGW,sorboGW}, where the present energy density per logarithmic $k$ interval of the gravitational waves relative to the critical density is given by
\begin{equation}
\Omega_{GW} h^2\simeq 3.5\times 10^{-7} \frac{H^2}{M_p^2} \left(1+4.3\times 10^{-7}\frac{H^2}{M_p^2} \frac{e^{4\pi\xi}}{\xi^6}\right).
\end{equation}
Here we evaluate $\Omega_{GW} h^2$ by taking $\xi=\xi(N)$ plotted in Fig.~\ref{phixi}, $H=H(N)$ given in Eq.~(\ref{Heom}), and $k=H_0 e^{N_0-N}$. When $N=N_0$, $k=H_0$ that corresponds to the present horizon of size $0.002\, {\rm Mpc}^{-1}$.  In Fig.~\ref{pbhgw}, we plot $\Omega_{GW} h^2$ against the frequency $f=3\times 10^{-18} e^{N_0-N} {\rm Hz}$ with $N_0=61$. The peak is associated with the production of about $14 M_\odot$ PBHs. The figure also shows the current upper limits on gravitational wave background made by the pulsar timing array experiments EPTA~\cite{EPTA},  NANOGrav~\cite{NANO}, and PPTA~\cite{PPTA}, and the projected sensitivity of SKA radio telescope~\cite{SKA}. 
The gravitational waves associated with the production of $14 M_\odot$ PBHs are about an order of magnitude below the current pulsar timing array sensitivity. Slightly lighter PBHs may source gravitational waves whose peak shifts to higher frequencies within the reach of the future SKA observation. Then, a full numerical calculation over a wide range of PBH masses should be carried out to assess the detectability or to constrain the formation of PBHs.

In a trapped inflation, the backreaction due to particle production damps the motion of the inflaton so that a successful inflation can be realized in a steep potential. This may solve the long standing problem in the slow-roll inflation that a flat potential is required. Moreover, the curvature perturbation induced by the particle production has distinctive non-Gaussianities~\cite{NGGW,meerburg}. In the case of a quadratic inflation, the measurements of CMB power spectra and bispectrum has constrained the inflaton-photon coupling $\alpha<32$~\cite{NGGW,meerburg}.  Furthermore, using $\alpha<23$, the formation of PBHs with mass $M_{\rm BH} < 10^8$g due to the photon production at the end of inflation has been discussed~\cite{lin,linde,cheng,bugaev}. Here we have used a combination of a flat and a steeper trapping potential with a smaller $\alpha=17$, leading to the production of high stellar-mass PBHs. These PBHs could be the cold dark matter observed by the LIGO detectors through a binary BH merger. We thus emphasize that PBHs are naturally produced in inflation with a trapping potential, although the PBH mass range depends on model parameters. Future observations of non-Gaussianities in CMB and large scale structures as well as direct measurements of gravitational wave background will be very important for testing the trapped inflationary scenario. 

This work was supported in part by the Ministry of Science and Technology, Taiwan, ROC under Grants No. MOST104-2112-M-001-039-MY3 (K.W.N.) and No. MOST104-2112-M-003-013 (W.L.).

\end{document}